\documentclass[aps,prl,superscriptaddress,preprint]{revtex4-2}

\usepackage{graphicx,amssymb,amsmath}
\usepackage[]{color}
\usepackage{ulem}

\begin{document}


\title{Experimental confirmation of the magnetic ordering transition induced by an electronic structure change in the metallic triangular antiferromagnet Co$_{1/3}$TaS$_2$}


\author{Han-Jin Noh}
\email{ffnhj@jnu.ac.kr}
\author{En-Jin Cho}
\affiliation{Department of Physics, Chonnam National University, Gwangju 61186, Republic of Korea}
\author{Byeong-Gyu Park}
\affiliation{Pohang Accelerator Laboratory, Pohang University of Science and Technology, Pohang 37673, Republic of Korea}
\author{Hyowon Park}
\email{hyowon@uic.edu}
\affiliation{Materials Science Division, Argonne National Laboratory, Argonne, Illinois 60439, USA}
\affiliation{Department of Physics, University of Illinois at Chicago, Chicago, Illinois 60607, USA}
\author{Ivar Martin}
\affiliation{Materials Science Division, Argonne National Laboratory, Argonne, Illinois 60439, USA}
\author{Cristian D. Batista}
\affiliation{Department of Physics and Astronomy, University of Tennessee, Knoxville, Tennessee 37996, USA}
\affiliation{Quantum Condensed Matter Division and Shull-Wollan Center, Oak Ridge National Laboratory, Oak Ridge, Tennessee 37831, USA}
\author{Pyeongjae Park}
\affiliation{Center for Quantum Materials, Seoul National University, Seoul 08826, Republic of Korea}
\affiliation{Department of Physics \& Astronomy, Seoul National University, Seoul 08826, Republic of Korea}
\affiliation{Materials Science and Technology Division, Oak Ridge National Laboratory, Oak Ridge, Tennessee 37831, USA}
\author{Woonghee Cho}
\affiliation{Center for Quantum Materials, Seoul National University, Seoul 08826, Republic of Korea}
\affiliation{Department of Physics \& Astronomy, Seoul National University, Seoul 08826, Republic of Korea}
\author{Je-Geun Park}
\affiliation{Center for Quantum Materials, Seoul National University, Seoul 08826, Republic of Korea}
\affiliation{Department of Physics \& Astronomy, Seoul National University, Seoul 08826, Republic of Korea}
\affiliation{Institute of Applied Physics, Seoul National University, Seoul 08826, Republic of Korea}


\date{\today}

\begin{abstract}
 
We report ARPES studies combined with DFT+DMFT calculations to confirm that the magnetic ordering vector transition from \textbf{Q}=(1/2,0,0) to \textbf{Q}=(1/3,0,0)  in the metallic triangular antiferromagnets Co$_{1/3\pm\epsilon}$TaS$_2$ ($\epsilon\approx$0.007) is induced by the electronic structure change in the system.
The ARPES-measured Fermi surface (FS) maps of Co$_{0.325}$TaS$_2$ show two hexagonal and one circular hole-like FSs around $\Gamma$, which matches well with the triple-\textbf{Q} state by taking into account the contribution of nesting vectors occurring between Co 3$d$ and Ta 5$d$ orbitals.
In the case of Co$_{0.340}$TaS$_2$, a new electron pocket around K appears and the FS geometry changes as a result of the correlation effect of Co$_4$S$_{18}$ tripods forming in the system.
The magnetic susceptibility calculations based on the charge-self-consistent DFT+DMFT band structures and the random phase approximation indicate that the most stable magnetic ordering vector (1/2,0,0) split into (1/6,0,0) and (1/2,0,0), which is consistent with the magnetic phase transition around $x$=1/3 in Co$_{x}$TaS$_2$.

\end{abstract}



\maketitle

\newpage
\section{Introduction}

Investigating how magnetic order interacts with electronic band structure has provided a major route to new physical phenomena and still represents a large part of unexplored territory in condensed matter physics.  
In particular, when the band structure has conduction electrons forming Fermi surfaces (FSs), the abundant passages of interaction have given an opportunity to novel quantum mechanical exotic states of matter and related phenomena.
Heavy fermion physics in Kondo lattices\cite{Coleman_HF, Yanagisawa}, colossal magnetoresistance in manganites\cite{Tokura_CMR},  topological Kondo insulators\cite{Coleman_TKI}, and quantum anomalous Hall effect (QAHE) in magnetic topological insulators\cite{Tokura_MTI}  can be exemplified.
In a broader sense, high T$_c$ superconductivity in cuprates\cite{Keimer} or iron pnictides\cite{Paglione} and Andreev reflection\cite{deJong} also fall into this category.
This approach has become even more powerful with remarkable advances both in band calculational methods\cite{DMFT} and angle-resolved photoemission spectroscopy (ARPES).\cite{ARPES}

Recently, a novel case study was reported implying that the interaction between magnetic order and electronic structure plays a crucial role in stabilizing an exotic magnetic ground state of matter, four-sublattice triple-\textbf{Q} (3\textbf{Q}) chiral ordering in a metallic triangular lattice antiferromagnet (TLAF) Co$_{1/3}$TaS$_2$.\cite{PPark_TripleQ}
It had been long sought in bulk crystals after theoretically predicted a decade ago.\cite{Martin1, Kato, Takagi}
This ordering has a special intriguing point in that it can be regarded as the short wavelength limit of magnetic skyrmion crystals corresponding to an effective magnetic flux quantum.\cite{Wang}
Once the ordering was experimentally realized in a bulk crystal, the immediate subsequent question is how can we manifest experimentally the details of the interaction between the electronic structure and the 3\textbf{Q} magnetic order.
This kind of study is quite rare in spite of its potential importance to quantum magnetism research due to the difficulty in controlling and isolating the effect of the electronic structure change only to the magnetic order.\cite{HPark}
In this perspective, the Co$_x$TaS$_2$ system provides a valuable chance to study the interaction in detail because the magnetic ground state is sensitively dependent on the Co content around $x$=1/3.\cite{PPark_PRB}
Thus, a fully comprehensive experimental study of the TLAF Co$_{1/3\pm\epsilon}$TaS$_2$ ($\epsilon\ll$1) can provide crucial information on microscopically dominant interactions and stabilizing mechanism for the 3\textbf{Q} state and the related exotic phases.

In this paper, employing synchrotron-based ARPES combined with density functional theory plus dynamic mean field theory (DFT+DMFT), we closely scrutinized the changes in the electronic structure of Co$_{1/3\pm\epsilon}$TaS$_2$ ($\epsilon\approx$0.007) single crystals having a minute difference in cobalt content to understand the origin of the magnetic ordering vector transition and the details for stabilization of the 3\textbf{Q} magnetic ordering.
The ARPES measurements reveal that the Fermi surface geometry changes drastically in spite of the small deviation from 1/3 moles of cobalt ions per formula unit.
It is definitely beyond electron filling effect in the rigid band picture, and can be explained only as a result of correlation effect induced by the slightly over-doped cobalt ions in the Co$_{0.340}$TaS$_2$ samples.
The magnetic susceptibility calculations based on the DFT+DMFT band structures and the random phase approximation (RPA) suggest that the FS geometry change is closely connected with the magnetic ordering vector transition  from \textbf{Q}=(1/2,0,0) to \textbf{Q}=(1/3,0,0) in reciprocal lattice units even though our DMFT spin susceptibility calculation does not include the momentum-dependent effects of self-energy and vertex function.
They also imply that the scatterings between Co 3$d$ and Ta 5$d$ electrons are the key process for the 3\textbf{Q} ordering stabilization.
These results demonstrate that although it is evident in theory but hard to verify experimentally, a local spin ordering transition can be induced solely through changes in the electronic structure of a metallic magnet.

\section{Experimental Details}
High-quality single crystal Co$_x$TaS$_2$ is prepared by the two-step chemical vapour transport (CVT) method following the recipe given in Ref.~\cite{PPark_PRB}, and parts of the single crystals in this work are the same batch samples as those used in Ref.~\cite{PPark_PRB}.
A shiny hexagonal shape single crystal is obtained, and each crystal was characterised by a superconducting quantum interference device magnetometer, MPMS-XL5 (Quantum Design, USA), along with electrical transport measurement by standard four-probe method with CFMS (Cryogenics. Ltd, UK).
The ARPES measurements were performed at the 4A1 beamline of the Pohang Light Source with a Scienta R4000 electron spectrometer and $\hbar\omega$=40$\sim$110 eV photons\cite{HDKim}.
The crystals were cleaved {\it in situ} by a top-post method at 25~K under $\sim3.0\times 10^{-11}$ Torr.
The total energy (momentum) resolution of ARPES data is $\sim$30~meV ($\sim$0.01~\AA$^{-1}$).
For the polarization dependent ARPES measurements, the experimental geometry was fixed, and the photon polarization was changed by the polarizing undulators in the beamline.
As to the $k_z$ dispersion of the ARPES data, we did not assign a specific value due to a low energy resolution in the $k_z$ direction and the quasi-two-dimensional structure of Co$_x$TaS$_2$, but treated them as projected, so the dispersion of the bands along the $k_z$ direction is neglected.

\section{Calculational Methods}
We adopt density functional theory plus dynamical mean field theory (DFT+DMFT) to study the correlated band structure and the magnetic susceptibility $\chi(q)$ of Co$_{1/3}$TaS$_2$. 
To perform DFT+DMFT, we first obtain the non-spin-polarized band structure using the Vienna Ab-initio Simulation Package (VASP) code. 
We used the experimental crystal structure of Co$_{1/3}$TaS$_2$, i.e. the $\sqrt{3} \times \sqrt{3}$ ordered phase by Co intercalants.
The Perdew-Burke-Ernzerhof (PBE) functional is used for the exchange-correlation energy of DFT. 
In DFT, we also used a 14$\times$14$\times$4 $k$-mesh with the energy cutoff of 400 eV for the plane-wave basis. 
Using the DFT non-spin-polarized band structure, we construct the tight-binding Hamiltonian that can reproduce effectively the same band structure to the DFT calculations, using the maximally localized Wannier functions for Co 3$d$ and Ta 5$d_{z^2}$ orbitals.
Since other Ta 5$d$ orbitals are located above the Fermi level, including them in the tight binding Hamiltonian is not essential.
This Hamiltonian is solved using DMFT by adopting the continuous-time quantum Monte Carlo (CTQMC) method as the impurity solver. 
We use the Hubbard interaction $U$=3 eV, the Hund’s coupling $J$=0.4 eV for Co 3$d$ orbitals, and the temperature T=116 K.
While the adopted $U$ value is smaller than the reported value obtained by the contrained-LDA method, it is within a reasonable range.\cite{Tesch}
We achieved the charge-self-consistency of DFT+DMFT using the DMFTwDFT code.\cite{Singh}
In DMFT, we used a fine $k$-mesh of 30$\times$30$\times$10. 
For the double counting potential, we used the fully-localized-limit formula of the double counting potential. 
The number of valence electrons is set to be 24.0. 
To impose the doping effect, we change the number of valence electrons and rerun the DMFT calculation for the fixed number of valence electrons.
To compute the magnetic susceptibility, we use the following formula based on the DMFT Green’s function, as explained in Ref.\cite{HPark}. 
First, we compute the polarizability $\chi^{0}(q)$ as follows: 
\begin{equation}
Re \chi^{0}(q) = \frac{1}{\pi}\int d\nu f(\nu) \sum_{k} \{Im G_{\alpha\beta}(k,\nu) \cdot Re G_{\beta\alpha}(k+q, \nu) + Re G_{\alpha\beta}(k,\nu) \cdot Im G_{\beta\alpha}(k+q,\nu)\}, 
\end{equation}
where $f(\nu)$ is the Fermi function and  $G_{\alpha\beta}$ is the DMFT Green’s function represented using the basis of the orbital character $\alpha$ and $\beta$. 
We use the maximum entropy method to obtain the DMFT self-energy on the real frequency.
If the DMFT self-energy is set to be zero, the $\chi^{0}(q)$ formula can be given as the Lindhard formula for the susceptibility. 
Therefore, it is expected that $\chi^{0}(q)$ will be strongly enhanced near the Fermi surface nesting vector, which can be computed using DMFT in this case. 
Then we compute the full magnetic susceptibility using the RPA formula:
\begin{equation}\chi(q) = \chi^{0}(q) + \chi^{0}(q) \ast \overline{U} \ast \chi(q),\end{equation}
where $\overline{U}$ is the RPA-type interaction between different orbitals. 
Here, we include both the intra-site interaction between Co $d$ orbitals and the inter-site interaction between Co $d$ and Ta $d$ orbitals.

\begin{figure}
\includegraphics[width=16.0 cm]{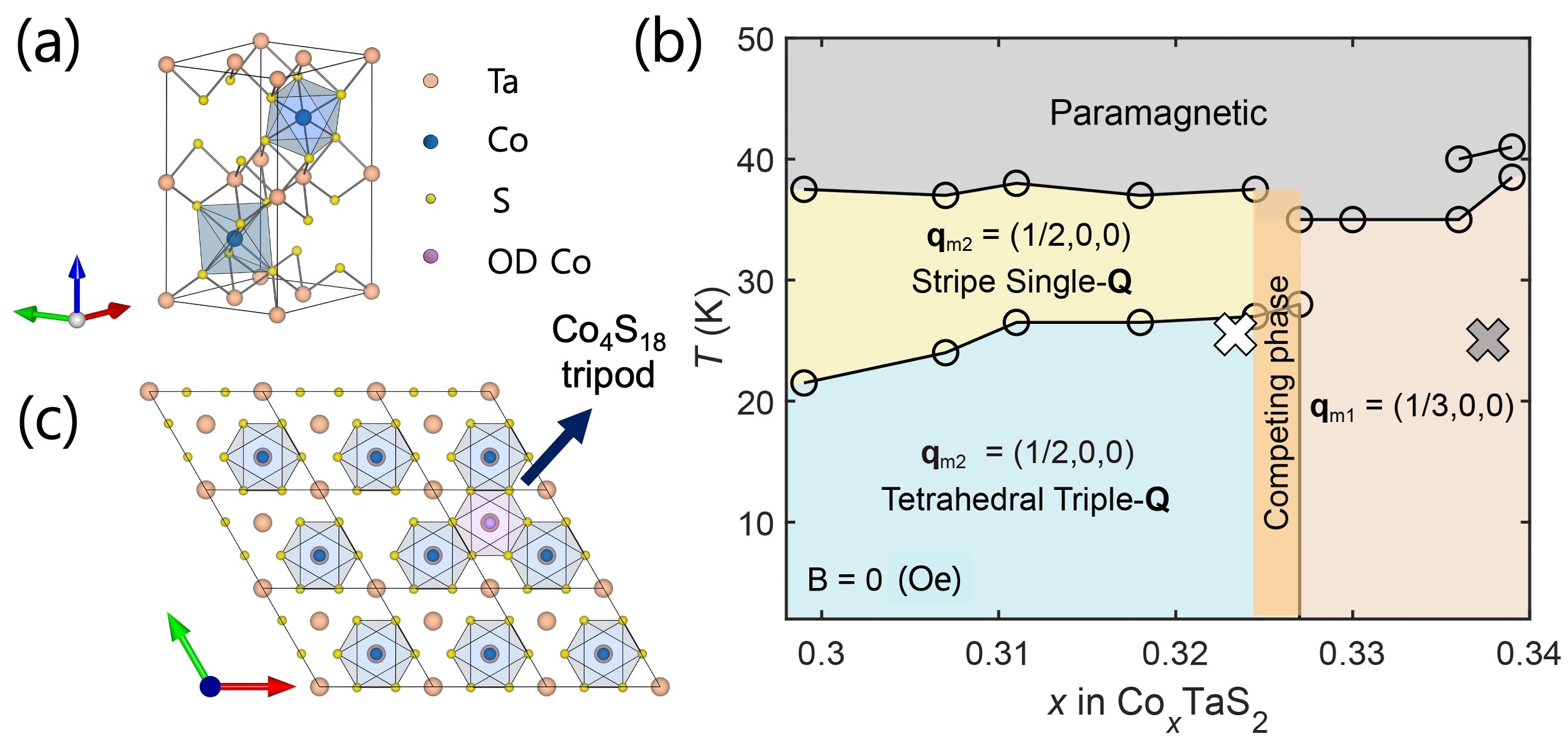}
\caption{\label{fig1}
\textbf{Crystal structure and magnetic phase diagram of Co$_x$TaS$_2$}
(a) The unit cell of Co$_{1/3}$TaS$_2$. (b) Composition-dependent magnetic phase diagram for Co$_{x}$TaS$_2$ (0.3 $< x <$ 0.34). \cite{PPark_PRB} The mark X denotes the Co composition in this work.
(c) A Co$_4$S$_{18}$ tripod in a van der Waals gap of over-doped Co$_{x}$TaS$_2$.
}
\end{figure}

\section{Results}

Co$_{x}$TaS$_2$ is Co-intercalated 2$H$-TaS$_2$, where the magnetic Co ions form a $\sqrt{3}\times\sqrt{3}$ ordered phase in the vdW gap when $x$=1/3.\cite{Parkin1}
The unit cell of the crystal structure is shown in Fig.~\ref{fig1}(a), in which a Co ion is located at the center of a S$_6$ octahedron with D$_{3h}$ symmetry.
The basic transport properties were measured in the 1980s to show a metallic behavior with an antiferromagnetic transition at 35 K, and a strong magnetic phase dependent Hall coefficient.\cite{Parkin2, Parkin3}
However, the recent revisited measurements on the single crystals revealed that the system with 0.30 $< x <$ 0.325 has another magnetic phase below 26.5 K to show a significant anomalous Hall effect (AHE), which is strongly coupled to the weak ferromagnetism.\cite{PPark_npjQM, PPark_PRB, JKim_ncom}
The exact magnetic structure for this phase was recently solved only after combining powder/single crystal neutron diffraction and AHE measurements, giving the tetrahedral 3\textbf{Q} ordering with \textbf{Q}=(1/2,0,0) in reciprocal lattice units.\cite{PPark_TripleQ}
The 3\textbf{Q} state with the AHE shows a sensitive dependence on the Co content $x$ as shown in the magnetic phase diagram  of Fig.~\ref{fig1}(b).
For the detailed spin configurations for the each magnetic phase, see Figs. 4(f)-(h) in Ref.~\cite{PPark_PRB}.
This sensitivity provides a rare opportunity to investigate what stabilizes the 3\textbf{Q} state and how it does so.
Although our previous ARPES measurements give a hint to the electronic origin for the magnetic transition, a close examination of combined experiment and theory for the electronic structure on both sides of the phase boundary can give the quantitative understanding for the stabilization mechanism.
Two different Co composition samples, $x$=0.325 (hereafter, UD-CTS) and 0.340 (OD-CTS), were chosen to measure the electronic structure by ARPES.
The Co content difference is just 0.015 moles per formula unit, so effectively there is no difference in Co ions arrangement except the existence of Co$_4$S$_{18}$ tripods in OD-CTS as shown in Fig.~\ref{fig1}(c).
The density of the tripods in this work is smaller than 0.7 \% per formula unit.

\begin{figure}
\includegraphics[width=16.0 cm]{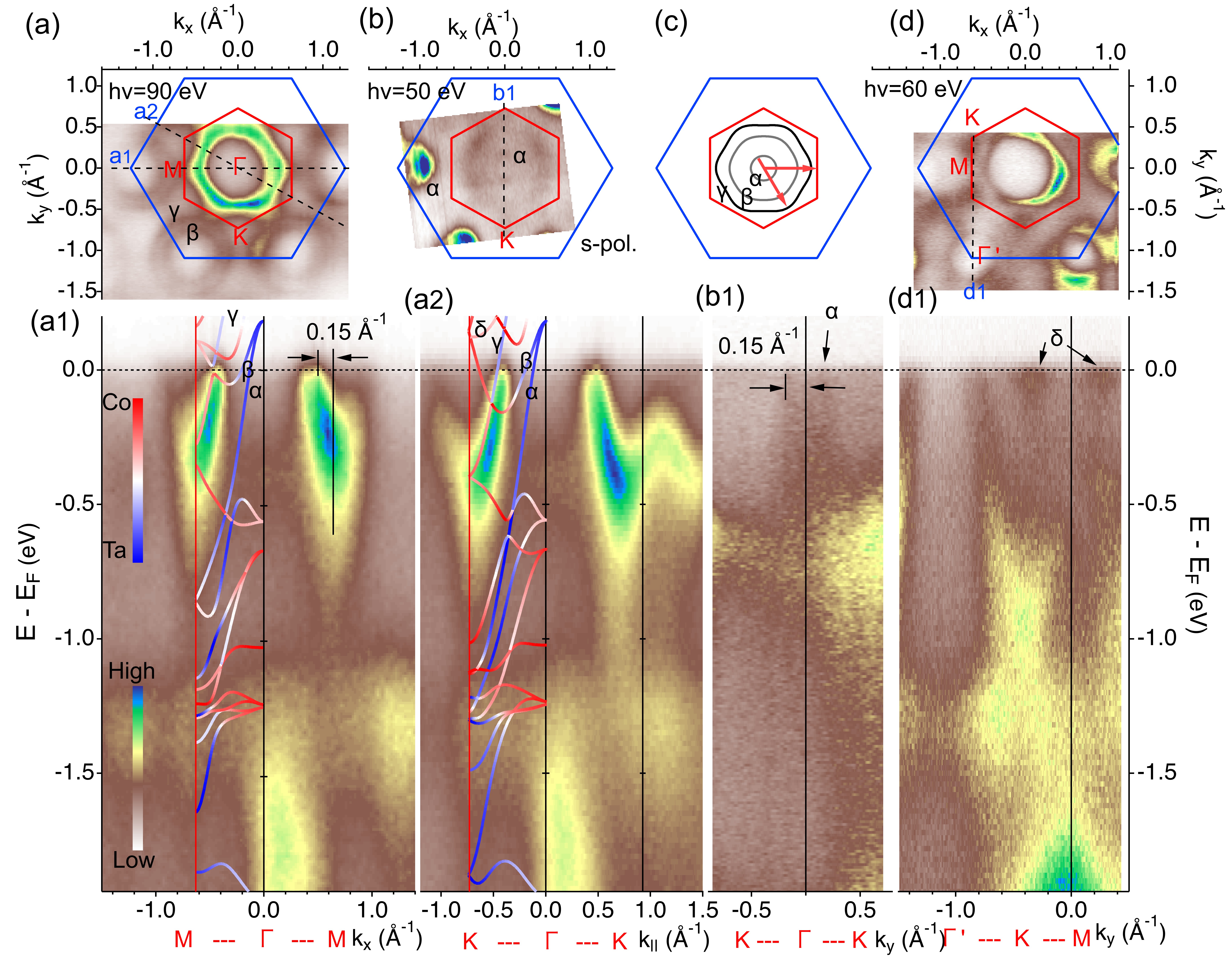}
\caption{\label{fig2}
\textbf{ARPES-measured electronic structure of Co$_{0.325}$TaS$_2$ }
(a) Measured FS map obtained with 90 eV photons. 
(a1, a2) The energy dispersion images with the Wannier function based DFT calculations along the dotted line a$_1$ and a$_2$ in (a). 
(b) FS map obtained with 50 eV photons in $s$-polarization mode. 
(b1) The energy dispersion image along the dotted line b$_1$ in (b).
(c) Schematic FS based on (a) and (b).
(d) FS map obtained with 60 eV photons.
(d1) The energy dispersion image along the dotted line d$_1$ in (d).
}
\end{figure}

The main features of the electronic structure for UD-CTS measured by ARPES are displayed in Figs.~\ref{fig2}(a)-(d).
The ARPES-measured Fermi surface (FS) map using $\hbar\omega$=90 eV photons at 25 K in Fig.~\ref{fig2}(a) looks very similar to the previously reported one for 2$H$-TaS$_2$ where a hexagonal double-walled hole pocket appears around $\Gamma$, K and K', respectively.\cite{Chatterjee}
These six FSs in a Brillouin zone (BZ) correspond to six Ta$^{4+}$ (5$d^{1}$) cations in a primitive unit cell in Fig.~\ref{fig1}(a).
In the case of the $\sqrt{3}\times\sqrt{3}$ ordering in Co$_{1/3}$TaS$_2$, the original BZ corresponding to 2$H$-TaS$_2$ denoted with the blue hexagon is reduced by 1/3 as denoted with the small red hexagon, so the K and K' points in the unfolded BZ become the second $\Gamma$ points in the folded zone.
As a result of this folding, the three double-walled hole pockets are also folded and redistributed by the interaction with each other, forming a conduction band structure with six hole-like pockets around each $\Gamma$ point equivalently.
In addition, the Co intercalants act as divalent cations with a localized spin $S$=3/2, raising the Fermi level of the conduction bands by adding 0.667 electrons per Ta cation.
This additional electron doping reduces the number of partially filled bands to three ($\alpha$, $\beta$, and $\gamma$) according to our Wannier function based DFT calculations as shown in Fig.~\ref{fig2}(a$_1$) and (a$_2$).

In the 90 eV ARPES data, only two FSs, $\beta$ and $\gamma$, are observed.
Also, the captured FSs in the first BZ and in the second BZs look a little different in shape and intensity.
This is due to the matrix effect in photo-ionization process, which is often useful to characterize conduction bands as is described below.
When we change the photon energy to 50 eV and the photon polarization perpendicular to the plane of incident (i.e. $s$-polarization), a small circular FS $\alpha$ around $\Gamma$ point appears as shown in Fig.~\ref{fig2}(b).
Interestingly, the hole pockets around the second $\Gamma$ points have high intensity while the one around the first $\Gamma$ point has weak intensity, showing a prototypical polarization and photon energy dependence.\cite{Noh_PdSb2}
For a more detailed comparison, the energy dispersion spectrum images along $\Gamma$-M direction and $\Gamma$-K direction are displayed with the DFT bands in Fig.~\ref{fig2}(a$_1$) and (a$_2$), respectively.
The overall structure is consistent with each other, but a considerable difference appears in the dispersion of $\beta$ around K.
The orbital character of the conduction bands is presented with a blue (Ta 5$d$) - red (Co 3$d$) color scale in the DFT bands, indicating the main character of each conduction band is Ta 5$d$ near the $\Gamma$ point but Co 3$d$ near zone boundary.
This orbital character distribution well explains why the $\alpha$ bands appear only in 50 eV photon energy data.
Since the photo-ionization cross-section ratio of Ta 5$d$ to Co 3$d$ at 50 eV is about ten times larger than that of 90 eV, the bands with Ta 5$d$ character have higher intensity in 50 eV data.\cite{Lindau}
Taking into account all the observed bands in different photon energies, the overall band structure measured by ARPES is consistent with that of DFT calculations.

Based on the measured FSs, a schematic FS map is obtained as displayed in Fig.~\ref{fig2}(c).
This FS structure is very favorable to the 3\textbf{Q} state with \textbf{Q}=(1/2,0,0) even though the stabilization mechanism is slightly modified.
In our previous work, the three quarters filling of the $\gamma$ band is ascribed to the major reason,\cite{PPark_TripleQ} but the more fine analysis in this work clearly reveals that there is a small gap of 0.15(1) \AA$^{-1}$ between the $\gamma$ band and M point as shown in Fig.~\ref{fig2}(a$_1$).
Also, the radius of $\alpha$ band FS is measured to be 0.15(1) \AA$^{-1}$ as shown in Fig.~\ref{fig2}(b$_1$).
Thus, this gap can be compensated if the inter-orbital scattering between the $\alpha$ and $\gamma$ bands is taken into account as is denoted with the red arrows in Fig.~\ref{fig2}(c).
The importance of this nesting effect is confirmed in the DMFT+DFT study as is described below. 

In addition to the three conduction bands, we observed an interesting extra structure around K in $\hbar\omega$=60 eV data as is shown in Fig.~\ref{fig2}(d).
When we see the energy dispersion relations of the structure along the dotted line d$_1$ in Fig.~\ref{fig2}(d), it looks like the tail of an electron pocket above the Fermi level.
The DFT calculation also indicates an unoccupied band $\delta$ at K just above the Fermi level as in Fig.~\ref{fig2}(a$_2$).
As will be explained later, the energy position of this band is one of the key factors in understanding the interaction between the magnetic ordering and the electronic structure in Co$_{1/3}$TaS$_2$.

\begin{figure}
\includegraphics[width=16.0 cm]{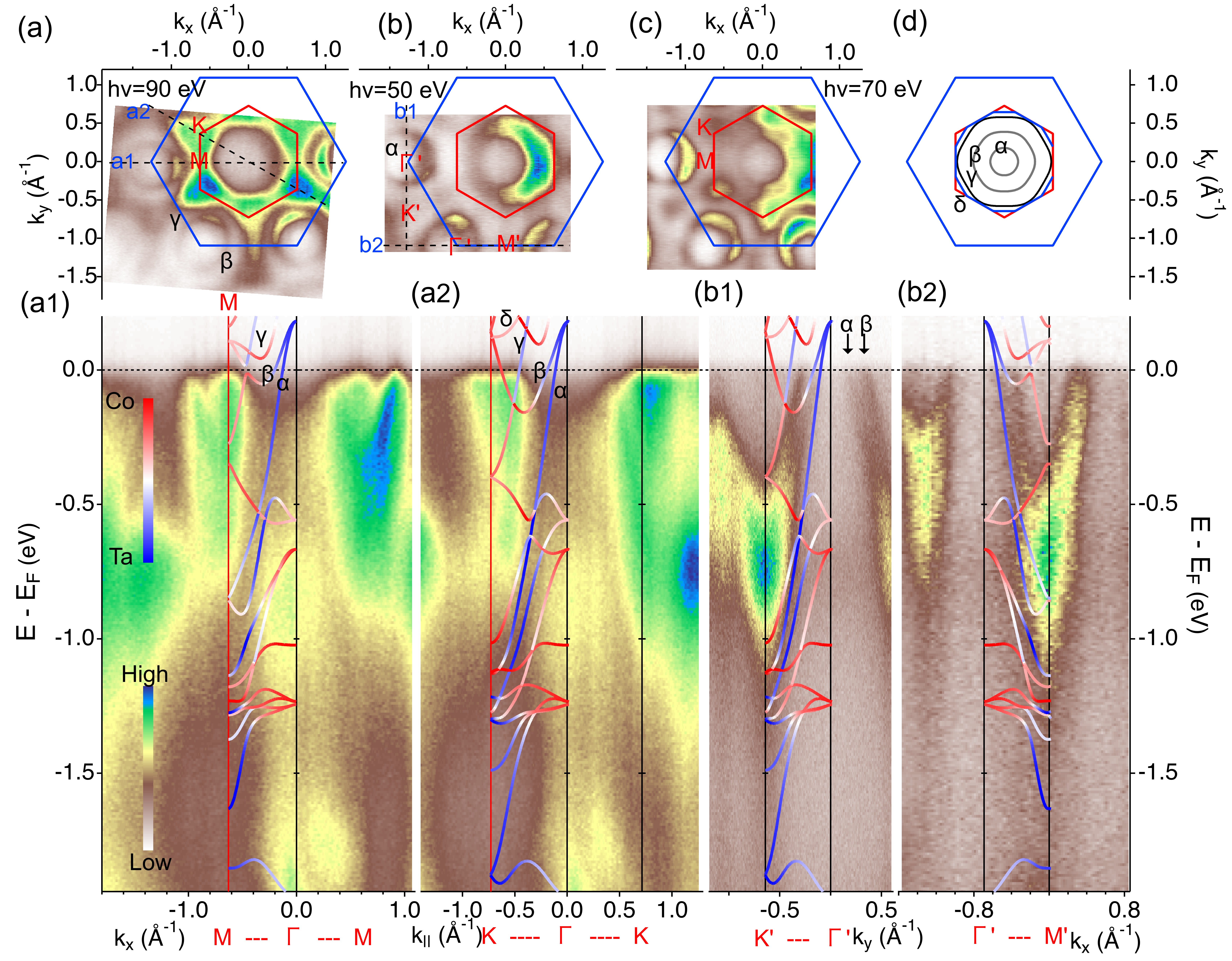}
\caption{\label{fig3}
\textbf{ARPES-measured electronic structure of Co$_{0.340}$TaS$_2$.}
(a) Measured FS obtained with 90 eV photons.
(a1, a2) The energy dispersion image along the dotted line a$_1$ and a$_2$ in (a).
(b) FS obtained with 50 eV photons.
(b1, b2) The energy dispersion image along the dotted line b$_1$ and b$_2$ in (b).
(c) Schematic FS based on (a) and (b).
(d) FS obtained with 70 eV photons.
}
\end{figure}

The electronic structure of UD-CTS is reasonably explained in terms of Co ordering phase in electron-doped 2$H$-TaS$_2$.
However, a small amount of additional Co intercalants to Co$_{1/3}$TaS$_2$ is found to induce a drastic change in the electronic structure.
In Figs.~\ref{fig3}(a)-(d), the main features of the ARPES-measured electronic structure for OD-CTS are displayed.
Just like the UD-CTS case, three conduction bands $\alpha$, $\beta$, and $\gamma$ are identified in the  $\hbar\omega$=90 eV data as is shown in Figs.~\ref{fig3}(a)-(a$_2$) and 50 eV data in Figs.\ref{fig3}(b)-(b$_2$), separately.
The most noticeable difference is that a large triangular electron pocket newly appears around K point as can be seen in Figs.~\ref{fig3}(a)-(a$_2$). 
Comparing the ARPES spectra with the DFT bands indicates that the band $\delta$ is the strongest candidate for the triangular electron pocket even though the band is above $E_F$ in the calculation.
Also, the orbital character of the electron pockets supports this assignment.
The DFT calculation confirms the band $\delta$ near K has mainly Co 3$d$ characters.
The measured intensity of the electron pockets shows a consistent result with this character-dependent photoionization cross-section ratio.
In the 90 eV data, it has the highest intensity, but it goes down in the 70 eV data as is seen in Fig.~\ref{fig3}(c), and then it almost disappear in the 50 eV data, which is parallel to the cross-section ratio.\cite{Lindau}
Based on the measured FSs, a schematic FS map is obtained as is displayed in Fig.~\ref{fig3}(d).
In this case, it is not clear whether the ordering vector \textbf{Q}=(1/2,0,0) is the most favorable because the triangular electron pockets change the majority portion of the nesting vectors in direction and size.

\begin{figure}
\includegraphics[width=10.0 cm]{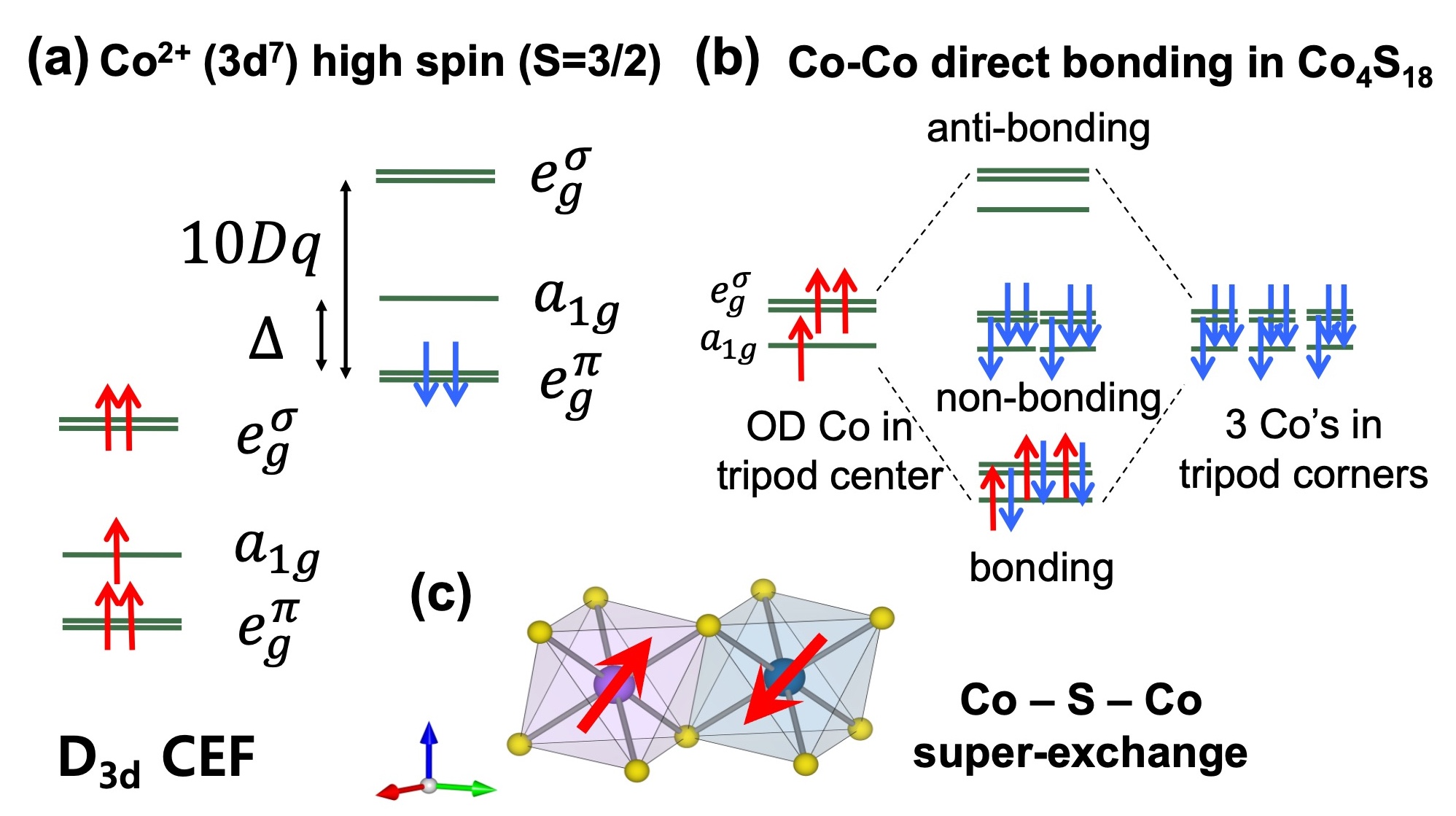}
\caption{\label{fig4}
\textbf{Schematic energy diagrams for Co$_4$S$_{18}$ tripod}
(a) Energy level splitting of Co 3$d$ orbitals under a crystal electric field of D$_{3h}$ symmetry.
(b) Schematic electron configuration for Co 3$d$-3$d$ direct bonding in a tripod. (c) A part of Co$_4$S$_{18}$ tripod. Edge sharing of CoS$_6$ octahedra enables not only the Co-Co direct bonding but also Co-S-Co super-exchange interaction.
}
\end{figure}

The drastic change of the electronic structure from UD-CTS to OD-CTS induced by the small amount of additional Co intercalants is definitely beyond a simple Fermi level tuning effect, and we argue that it is qualitatively explained by introducing electron correlation effects mediated by Co$_4$S$_{18}$ tripods in OD-CTS.
An additional Co intercalant more than 1/3 moles per formula unit occupies a vacant Co site with $z$=c/4 or 3c/4 (the 2$d$ Wyckoff position in the space group $P6_322$) in a Co layer as illustrated in Fig.~\ref{fig1}(c), which forms a Co$_4$S$_{18}$ tripod.\cite{PPark_npjQM}
Once the system hosts Co$_4$S$_{18}$ tripods, they are expected to induce additional interaction channels in the electronic structure and in the magnetic structure, respectively.
Since a Co atom occupies the center of a trigonal anti-prismatic S$_6$ octahedron with $D_{3d}$ symmetry, the Co 3$d$ orbitals are split into three manifolds e$^{\pi}_{g}$, a$_{1g}$, and e$^{\sigma}_{g}$ as shown in Fig.~\ref{fig4}(a).
A simple perturbative estimate of the crystal-field energy splitting, based on the atomic positions obtained from diffraction refinements,\cite{PPark_npjQM} yields an energy difference $\Delta$ between a$_{1g}$ and e$^{\pi}_{g}$ states of approximately 0.31 of the total crystal field splitting $10Dq$.
Then, a divalent Co cat-ion with seven 3$d$ electrons has half-filled e$^{\sigma}_{g}$ and a$_{1g}$ states with a high spin configuration $S$=3/2.
The distance between the nearest Co ions in a tripod is 3.31 \AA, which is the second shortest bonding in the system.\cite{PPark_PRB}
At this distance, direct overlap between Co 3$d$ orbitals is non-negligible.
Within a tripod, the central Co 3$d$ orbitals hybridize equally with those of the three corner Co atoms, with a coupling strength $V_{dd}$.
Because hybridization among the corner Co sites is negligible, the Co cluster in the tripod forms one bonding, two non-bonding, and one anti-bonding state per 3$d$ orbital, resulting in an energy gain of $\sqrt{3}V_{dd}$.
If the bonding energy gain $\sqrt{3}V_{dd}$ exceeds the on-site Coulomb energy cost in e$^{\sigma}_{g}$ or a$_{1g}$ states, electrons in the e$^{\sigma}_{g}$ or $a_{1g}$ levels form spin-singlet bonding states, as illustrated in Fig.~\ref{fig4}(b).
In the band structure, this direct bonding leads to a down-shift of Co character bands, which explains the appearance of the $\delta$ band electron pocket at the K point.
This argument is also consistent with the observation that the change in the electronic structure appears dominantly only in the $\delta$ bands.
Since these are derived from Co $d$-$d$ interactions, bands with strong Co 3$d$ characters are effectively affected.
As is shown in Fig.~\ref{fig3}(a1), all conduction bands exhibit enhanced Co contributions near M and K points, while the Co 3$d$ weight is reduced near $\Gamma$ point.
Because the conduction band minimum of the $\delta$ band is located at K, it is qualitatively parallel with the observed behavior.

From the magnetic structural viewpoint, a new magnetic exchange channel appears in a Co$_4$S$_{18}$ tripod.
In a Co under-doped regime without a tripod, the localized spins in Co$^{2+}$ ions interact with each other mainly via conduction electrons due to the long distance between the Co ions.
Meanwhile, in a Co over-doped regime with tripods, the Co-S-Co super-exchange interaction additionally becomes available between the Co$^{2+}$ ions in the tripods as is schematically shown in Fig.~\ref{fig4}(c).
Since the super-exchange favors antiferromagnetic interaction, OD-CTS is expected to show stronger antiferromagnetic behaviors than UD-CTS.
However, when we consider the fact that the density of tripods in OD-CTS is below 2\% and that the 3\textbf{Q} state also has an antiferromagnetic exchange energy between the localized spins of Co$^{2+}$ ions, a small exchange energy change induced by the super-exchange interaction in a tripod can hardly be a dominant factor to the magnetic ordering vector transition from \textbf{Q}=(1/2,0,0) to \textbf{Q}=(1/3,0,0).

\begin{figure}
\includegraphics[width=18.0 cm]{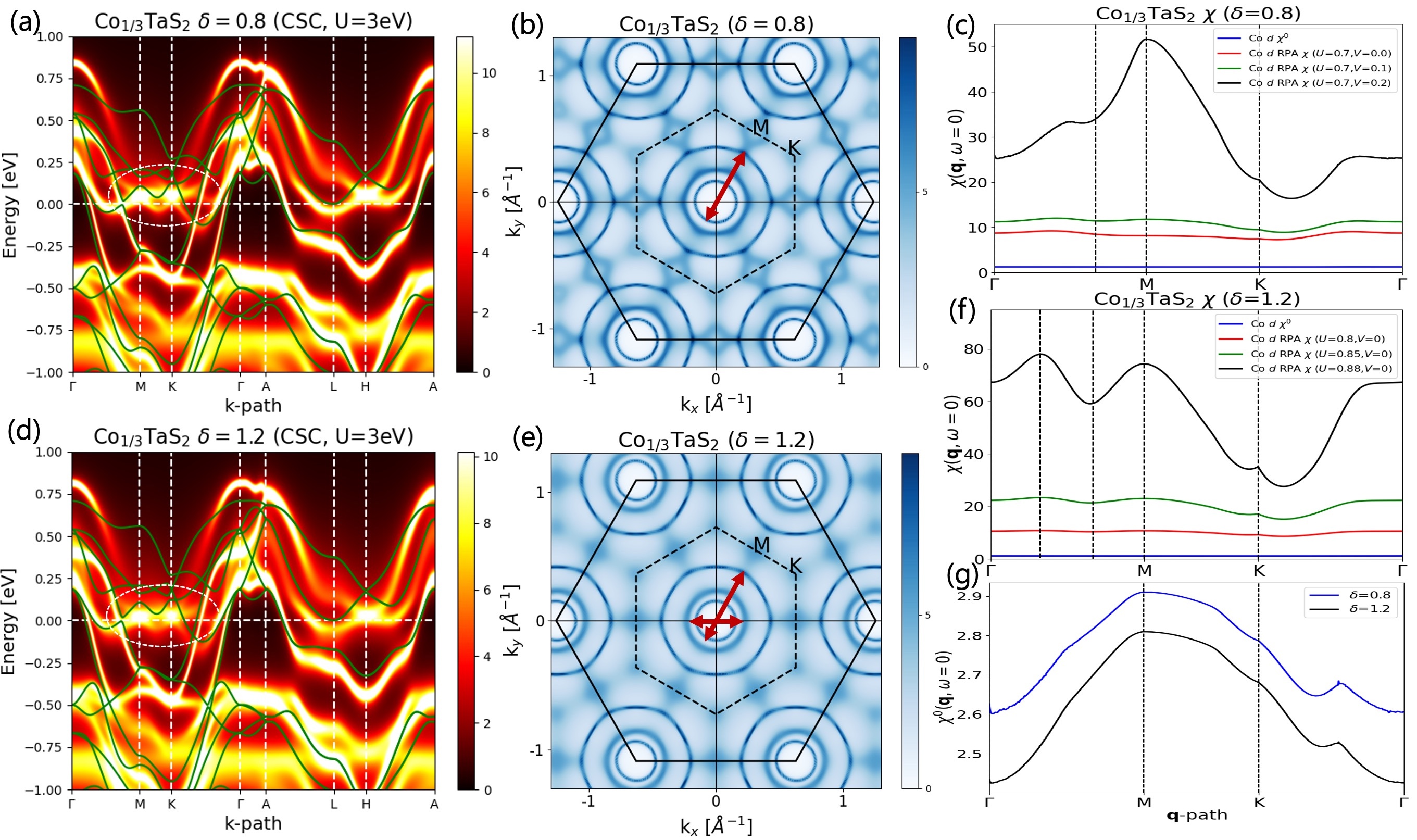}
\caption{\label{fig5}
\textbf{DMFT calculations for UD-CTS and OD-CTS}
(a) The DMFT band structure for UD-CTS. The green lines are the DFT bands.
(b) The DMFT FS for UD-CTS.
(c) The magnetic susceptibility for UD-CTS.
(d-f) The same calculations to (a-c) for OD-CTS. 
(g) The doping dependence of the polarizability $\chi^{0}(q,\omega=0)$.}
\end{figure}

In order to theoretically confirm the dominancy of the electronic structure effect to the spin ordering pattern, we performed the band structure calculations based on charge-self-consistent (CSC) density functional theory plus dynamical mean field theory (DFT+DMFT), and computed the magnetic susceptibility based on the DMFT band structures.
The detailed calculation methods are described in Calculational Methods.
Figures \ref{fig5}(a)-(c) display the DMFT band structure, the DMFT FS, and the magnetic susceptibility calculations for UD-CTS, respectively.
To simulate the band structure of UD-CTS, we use the Hubbard interaction $U$=3 eV, the Hund’s coupling $J$=0.4 eV in the calculation.
Even though the $U$ value is somewhat smaller than the typically used value in transition metal ions, it gives an optimal result for our ARPES-measured band structure of Co$_{x}$TaS$_2$ within a reasonable parameter range.\cite{HPark, Tesch}
The Fermi level was adjusted by introducing an additional doping parameter $\delta$=0.8, which means an additional number of valence electrons per crystallographic unit cell, i.e. per two Co cations.
This additional doping in DMFT is necessary to compare with the surface-sensitive ARPES measurements.
The DMFT band structure shows the renormalization of the DFT one (green line) near the Fermi energy.
If we compare the conduction bands in the DMFT FS with the experimental ones, the calculated $\gamma$ band is smaller than the experimental one. 
However, the major features of UD-CTS band structure are qualitatively well reproduced in the calculation.
The momentum dependency of the magnetic susceptibility $\chi(q)$ was also checked as shown in Fig.~\ref{fig5}(c) using the eq. (1) and (2) described in Calculational Methods.
 While the polarizability $\chi^{0}(q)$ has the weak momentum dependence, the full susceptibility shows the strong momentum dependence, which is enhanced by the increase of the RPA-type inter-orbital interaction $V$ between the Co 3$d$ and Ta 5$d$ orbitals.
For UD-CTS, the leading peak position is at M, indicating that the nesting of \textbf{Q}=(1/2,0,0) occurs between the Co 3$d$ ($\gamma$ band) and Ta 5$d$ ($\alpha$ band) orbital characters as denoted with red arrow in Fig.~\ref{fig5}(b).
For OD-CTS, we increase the additional doping parameter $\delta$ to 1.2, which moves parts of the band structure near K towards below the Fermi level.
This is clearly seen by comparing the calculated bands denoted with a white dotted ellipse in  Fig.~\ref{fig5}(a) and (d), respectively.
The difference $\Delta\delta$=0.4 in the calculations corresponds to additional electron doping of 0.2 per Co ion, while the electron doping difference between OD-CTS and UD-CTS in the experiments is $\sim$0.042 electrons per Co ion.
This large difference clearly indicates that the FS changes are not induced by a simple Fermi level shift in rigid bands, and electron correlation effects from the tripods have to be taken into account.
As is described above, the Co-Co direct bondings in the tripods play a key role to push down the energy states with relatively strong Co 3$d$ orbital character near the zone boundary including K point.
Unfortunately, the effect of the low density tripods and the short-range correlation between Co ions can not be fully incorporated in the DFT+DMFT scheme, so the deviation between the calculations and the ARPES data is inevitable to some degree.
For example, the spectral intensity near K in Fig.~\ref{fig5}(e) is much weaker than that of ARPES data.

In spite of the inevitable deviation, the calculated magnetic susceptibility explains the change of the magnetic ordering vector \textbf{Q}=(1/2,0,0) in the over-doped region as shown in Fig.~\ref{fig5}(f).
In the magnetic susceptibility calculations, it is worth mentioning that not only the electron doping but also the interplay between the FS nesting vector and the RPA-type interactions plays an important role to the transition of the magnetic ordering vector.
Although the change of $\chi^0(q)$ is not too sensitive to the doping, the nesting effect based on multiple peaks can be clearly seen as in Fig.~\ref{fig5}(g).
While the leading peak position is located at the M point due to the inter-band nesting effect of the Ta 5$d$ hole pockets (red arrow in Fig.~\ref{fig5}(b)), the second highest peak, which is possibly responsible for the helical order in the over-doped region, is located near \textbf{Q}=(1/3,0,0) due to the nesting effect within Co 3$d$ hole pockets (additional red arrow in Fig.~\ref{fig5}(e)).
Upon the electron doping, the peak near \textbf{Q$_{1/3}$}=(1/3,0,0) moves to \textbf{Q$_{1/6}$}=(1/6,0,0) and is more strongly enhanced than the other comparable peak at \textbf{Q$_{1/2}$}=(1/2,0,0) to become the leading peak at $\delta$=1.2 due to the large RPA-type intra-atomic Co 3$d$ interaction $\overline{U}$ as shown in Fig.~\ref{fig5}(f).
This definitely makes \textbf{Q$_{1/2}$} unstable and \textbf{Q$_{1/6}$} more stable for the magnetic ordering vector.
However, the peak weight at \textbf{Q$_{1/2}$} is still very comparable to the new leading peak at \textbf{Q$_{1/6}$}, so the resulting magnetic ordering vector is not simply predictable.
The experimentally determined ordering vector \textbf{Q$_{1/3}$} is probably a competing result between the two \textbf{Q} vectors.
Even though the exact leading peak position in OD-CTS is not clearly derived from the magnetic susceptibility calculations, this matches well with the nesting vector within the Co 3$d$ bands, indicating that the FS geometry dominates the magnetic ordering patterns in Co$_{1/3+\epsilon}$TaS$_2$.

\section{Conclusion}
In conclusion, we performed ARPES studies combined with CSC-DFT+DMFT calculations to confirm that the magnetic ordering vector transition in Co$_{1/3\pm\epsilon}$TaS$_2$ ($\epsilon\approx$0.007) is induced by an electronic structure change.
The ARPES-measured FS map of UD-CTS has two hexagonal and one circular hole-like Fermi surfaces around $\Gamma$, which matches well with the 3\textbf{Q} state by taking into account the contribution of nesting vectors occurring between Co 3$d$ and Ta 5$d$ orbitals.
In OD-CTS case, a new electron pocket around K appears and the FS geometry changes as a result of the correlation effect of Co$_4$S$_{18}$ tripods.
The magnetic susceptibility calculations based on the DFT+DMFT band structures and the RPA indicate that the most stable magnetic ordering vector \textbf{Q}=(1/2,0,0) split into \textbf{Q}=(1/6,0,0) and (1/2,0,0), which is consistent with the magnetic phase transition around $x$=1/3 in Co$_{x}$TaS$_2$.

\section{Acknowledgments}
This work was supported by Global-LAMP program of the National Research Foundation (NRF) of Korea Grant funded by the Ministry of Education (Grant No. RS-2024-00442775).
H. Park acknowledges the computing resources provided on Bebop, a high-performance computing cluster operated by the Laboratory Computing Resource Center at Argonne National Laboratory.
P. Park, H. Park, and I. Martin acknowledge support by the U.S. Department of Energy, Office of Science, Basic Energy Sciences, Materials Science and Engineering Division.
Work at Seoul National University was supported by the Samsung Science \& Technology Foundation (Grant No. SSTF-BA2101-05). 
J.-G. Park is partly funded by the Leading Researcher Program of the NRF of Korea (Grant No. 2020R1A3B2079375).



\end{document}